\documentclass{iaus}
\usepackage{graphicx}

\title[Secular Evolution]
{Density-Wave Induced Morphological Transformation
of Galaxies along the Hubble Sequence}

\author[Zhang and Buta]
{Xiaolei Zhang$^1$ and
Ronald J. Buta$^2$}

\affiliation{$^1$Department of Physics and Astronomy,
\break George Mason University, Fairfax, VA 22030, USA 
\break email: xzhang5@gmu.edu\\[\affilskip]
$^2$Department Physics and Astronomy, University of
Alabama, \break 514 University Blvd E, Box 870324,
Tuscaloosa, AL 35401, USA \break email: buta@sarah.astr.ua.edu}

\jname{Dynamics and Evolution of Disc Galaxies}
\editors{A. Zasov and D. Pfenniger eds.}
\begin{document}
\maketitle

\begin{abstract}
In the past two decades, secular evolution has emerged as an
important new paradigm for the formation and evolution of the
Hubble sequence of galaxies. A new dynamical mechanism
was identified through which density waves
in galaxies, in the forms of nonlinear and global spiral and bar 
{\em modes}, induce important collective dissipation effects 
previously unknown in traditional studies.  These effects lead to the
evolution of the {\it basic state} of the galactic disk, consistent
with the gradual transformation of a typical galaxy's morphological
type from a late to an early Hubble type.  In this paper,
we review the theoretical framework and highlight our recent result
which showed that there are significant qualitative and quantitative
differences between the secular evolution rates predicted by
the new theory compared with those predicted by the classical
approach of Lynden-Bell and Kalnajs.  These differences are the outward
manifestation of the dominant role played by collisionless shocks
in disk galaxies hosting quasi-stationary, extremely non-linear
density-wave modes.
\end{abstract}

\section{Introduction}

The possibility that galaxy morphologies can transform significantly
over their lifetime, not only through violent episodes such as merger
or satellite accretion, but also through slow but steady internal
secular dynamical processes, is a notion that is gaining acceptance in the
recent decades.  In the past, the work on secular evolution has been focused on
gas accretion in barred galaxies and the growth of pseudo bulges (Kormendy \&
Kennicutt 2004 and the references therein).  This is partly due to the
long-held notion that gas is the only mass component capable of dissipation,
and the stellar component is adiabatic and generally does not lose or gain
energy and angular momentum as they orbit around the center of a galaxy.

The first indication that there is the possibility for significant {\it stellar
mass redistribution} in galaxies originates from the seminal work of Lynden-Bell
and Kalnajs (1972, hereafter LBK), who showed that a trailing spiral density wave
possesses a gravitational torque that over time can transport angular
momentum outward.  LBK were interested in the angular momentum transport phenomenon
because they were seeking a generating mechanism for the spiral density waves, thought
to be short-lived wave trains constantly being amplified out of noise and
subsequently absorbed at the inner Lindblad resonance.  
Since the density wave is considered
to possess negative energy and angular momentum inside corotation relative
to the basic state (i.e. the axisymmetric disk), an outward angular momentum
transport would encourage the spontaneous growth of the wave trains.  LBK
at that time was not interested in the secular morphological evolution of
the basic state of the disks.  In fact, in the same paper, they showed that
for WKBJ (tightly wrapped) waves, the long-term energy and angular momentum
exchange between the wave and the basic state is zero away from
the wave-particle resonances.  This is possible in
the presence of the outward
angular momentum transport by gravitational torque couple because
they showed that there is a second type of torque couple, the so-called
advective torque couple (due to lorry transport), that opposes the gravitational
torque couple, and the sum of the two types of torque couples is a constant
independent of the galactic radii.  The total torque couple, which is equal 
to the rate of total radial angular momentum flux, is thus a constant during the
outward angular momentum
transport, and there is no interaction of the wave and basic state
except at the wave-particle resonances (i.e., they thought the wave picks
up angular momentum from the basic state at the inner Lindblad resonance
and dumps it at the outer Lindblad resonance, and en route of this radial
transport the total angular momentum flux remains constant).

Zhang (1996,1998,1999) showed that the classical theory of LBK ignored
an important collective dissipation process present in the gravitational
N-body disks possessing {\em self-organized, or spontaneously-formed, density
wave modes}. This process is mediated by collisionless shocks at the
density wave crest, which breaks the adiabaticity or the conservation
of the Jacobi integral condition -- a condition which is shown to be valid only
for a {\em passive} orbit under an {\em applied} spiral
or bar potential, and is now shown not to be obeyed by orbits
undergoing collective dissipation.  
The overall manifestation of the collective dissipation
process is an azimuthal phase shift between the potential and the density
distribution of the density wave pattern, and for a self-sustained mode
this phase shift is positive inside corotation, and negative outside.
The presence of the phase shift means that for every annulus of the galaxy, there
is a secular torque applied by the density wave on the disk matter in the
annulus, and the associated energy and angular momentum exchange between the wave
and the basic state of the disk.  As a result the disk matter inside corotation
(both stars and gas) loses energy and angular momentum to the wave, and
spirals inward, and the disk matter outside corotation gains energy
and angular momentum from the wave and spirals outward.  This energy
and angular momentum exchange between the wave and the basic state
of the disk thus becomes the ultimate driving mechanism for the secular
evolution of the mass distribution of the basic state of galaxy disks.
The energy and angular momentum received by the wave from the basic
state, incidentally, serve as a damping mechanism for the spontaneously
growing unstable mode, allowing it to achieve quasi-steady state
at sufficiently nonlinear amplitude.

In Zhang (1998), a set of analytical expressions for the secular
mass accretion/excretion rate was derived, and was confirmed quantitatively
in the N-body simulations presented in the same paper.  However, due
to the 2D nature of these simulations, where the bulge and halo were
assumed to be spherical and inert, the simulated wave has an average
density contrast of 20\% and potential contrast of 5\%, both much lower than the
average observed density wave contrast in physical galaxies, so the simulated
disk did not evolve a lot (Zhang 1999), despite the fact that these low
evolution rates conform exactly to the analytical formula's prediction 
for the corresponding wave amplitude (Zhang 1998).

Zhang \& Buta (2007) and Buta \& Zhang (2009)
used near-frared images of observed
galaxies to derive the radial distribution of the azimuthal 
potential-density phase shifts, and to use the positive-to-negative 
zero crossings of the phase shift curve to determine the corotation radii 
(CRs) for galaxies possessing spontaneously-formed density wave modes.
This approach works because the alternating
positive and negative humps of phase shift distribution lead to 
the correct sense of energy and angular momentum exchange between the wave mode
and the disk matter to encourage the spontaneous emergence of the mode.
In Zhang \& Buta (2007) and Buta \& Zhang (2009)
we have found good correspondence between the predicted CRs 
using the potential-density phase shift approach with the resonance
features present in galaxy images, and also with results from other
reliable CR determination methods within the range of validity of
these methods.  Beside CR determination, an initial
test case for mass flow rate calculation,
for galaxy NGC 1530, was also carried out in Zhang \& Buta (2007) using
the same volume-torque-integration/potential-density phase shift approach,
and there we found that since this galaxy has exceptionally large surface
density and density-wave arm-to-interarm contrast, mass flow rate
more than 100 solar mass per year were obtained across much of the 
galactic radii for this galaxy.  This level of mass flow rate is more
than sufficient to transform the Hubble type of a late type galaxy to
an early type within a Hubble time.  Other galaxies we have tested have
significantly lower mass flow rates but still are sufficient to
lead to significant mass redistribution over a Hubble time.

Recently, we have applied the potential-density phase shift/volume-torque 
method to a larger sample of galaxies in order to estimate their 
mass flow rates.  In related earlier works, Gnedin et al.  (1995) 
and Foyle et al. (2010) have applied LBK type gravitational
torque integral to the calculation of angular momentum redistribution
rate in a number of galaxies.  In our own studies, we found that such earlier work
using the gravitational torque couple alone had significantly
under-estimated the (implied) total mass flow rates in these galaxies.
As it turns out, the advective torque couple, which opposes the
gravitational torque couple in the LBK classical theory, becomes
to have the same sense of angular momentum transport direction for spontaneous-formed
density wave modes at the nonlinear regime.  Furthermore, at the extremely 
non-linear wave amplitudes usually found for observed galaxies,
the contribution of advective torques due to the collisionless
shocks far exceeds the contribution from gravitational torques,
and becomes the dominant driver for the secular evolution of
galaxy mass redistribution.  The sum of both types of (surface) torque couples
turns out to be equal to the (integral of the) volume-type torque we
used in this work, which is first proved in Zhang (1998, 1999).
The past calculations of gas mass accretion near the central region
of galaxies (e.g. Haan et al. 2009) are likely to have
significantly underestimated the gas mass flow rate for the same
reason.

Our current work also has other important implications for the
fundamental questions of galactic dynamics.  For example, on
the modal versus transient nature of the density wave patterns
in galaxies (see, e.g. Sellwood 2010 and the references therein).  
The bell-shaped total angular momentum flux or torque
coupling integral, which is equivalent to the two-humped phase shift or
volume torque distribution, that we have found to be overwhelmingly
present in observed galaxies, has no explanation in the classical
LBK theory of transient waves, which predicts constant total angular
momentum flux for a wave train between the inner and outer Lindblad
resonances (LBK; Binney \& Tremaine 2008), but is a natural consequence 
of the spontaneously-formed intrinsic modes
of disk galaxies, as first demonstrated in Zhang (1998).  Also,
the result of CR determination for the majority of the more 
than 150 galaxies analyzed in Buta \& Zhang (2009)
using the potential-density phase shift method
also supports the modal view, since for transient waves one should not be
able to use the Poisson equation alone to
predict a partially-kinematic quantity such as the corotation radius.
For the successful application of the potential-density phase shift method,
the Poisson equation and the equations of motion must have achieved a good 
degree of mutual
consistency to allow a quasi-steady state to form, 
a condition naturally met by self-sustained modes, and in general
not expected to hold for transient waves.

\section{THEORETICAL BASIS FOR APPLYING THE POTENTIAL-DENSITY PHASE SHIFT
APPROACH TO THE MASS FLOW RATE CALCULATION}

The detailed discussion on the
new dynamical mechanism responsible for the secular mass redistribution
in galaxies (both stellar and gaseous) is described in Zhang 1996,1998,1999).
We now briefly summarize the derivations relevant
to the calculation of the mass flow rate in galaxies.

The (inward) radial mass accretion rate at a galactic radius R can be written as
\begin{equation}
{dM \over dt} = - {dR \over dt} 2 \pi R \Sigma_0(R)
\end{equation}
where $\Sigma_0(R)$ is the mean surface density of the basic state
of the disk at radius R, and $-dR/dt$ is the mean orbital delay rate of
an average star.

We also know that the mean orbital decay rate of a single star is related to its
angular momentum loss rate $dL*/dt$ through
\begin{equation}
{dL^* \over dt} = - V_c M_* {dR \over dt}
\end{equation} 
where $V_c$ is the mean circular velocity at radius R, and $M_*$ the mass of the relevant star.

Now we have also
\begin{equation}
{dL^* \over dt} = \bar{dL \over dt} {M_*  \over \Sigma_0}
\end{equation} 
where $\bar{dL \over dt}$ is the angular momentum loss rate of the basic
state disk matter per unit area at radius R.

Since
\begin{equation}
\bar{dL \over dt} = { 1 \over {2 \pi}} \int_0^{2 \pi}
\Sigma_1 { {\partial {\cal V}_1} \over {\partial \phi}} d \phi
\end{equation}
(Zhang 1996), we have finally

\begin{equation}
{dM \over dt} = {R \over {V_c}} \int_0^{2 \pi} \Sigma_1  {{\partial {\cal V}_1} \over {\partial \phi}} d \phi
\end{equation}
where the subscript 1 denotes the perturbation quantities.

In the above derivation we have used a volume-type of torque $T_1(R)$,
\begin{equation}
T_1(R) \equiv R \int_{0}^{2 \pi} 
\Sigma_1  {{\partial {\cal V}_1} \over {\partial \phi}} d \phi,
\end{equation}
which was first introduced in the context of the self-torquing of the disk matter
by its associated spontaneously-formed density wave modes in Zhang [1996,1998]. 
The volume torque is equal to
the time rate of angular momentum exchange between the density wave and the
disk matter in a unit-width annulus 
located at galactic radius R, for wave modes in approximate
quasi-steady state.  In the past, two other types of torque-coupling integrals have
also been used (Lynden-Bell \& Kalnajs 1972; Binney \& Tremaine 2008).  These
are the gravitational torque couple $C_g(R)$
\begin{equation}
C_g(R) = {R \over {4 \pi G}} \int_{- \infty}^{\infty} \int_0^{2 \pi}
{{\partial {\cal V}} \over {\partial \phi}}
{{\partial {\cal V}} \over {\partial R}}
d \phi dz ,
\end{equation}
and the advective torque couple $C_a(R)$
\begin{equation}
C_a(R) = R^2 \int_0^{2 \pi} \Sigma_0 V_R V_{\phi} d \phi ,
\end{equation}
where $V_R$ and $V_{\phi}$ are the radial and azimuthal velocity perturbation
relative to the circular velocity, respectively.

In the classical theory, the volume torque integral $T_1(R)$
can be shown to be equal to $dC_g/R$ in the linear regime
(Zhang 1998, original derivation due to S. Tremaine, private communication).
However, for spontaneously-formed density wave modes,
when the wave amplitude is significantly nonlinear and the
importance of collisionless shocks at the density wave crest
begins to dominate, it can be shown that one of the
crucial conditions in the proof of the $T_1(R) = dC_g/dR$ relation,
that of the validity of the differential form of the Poisson
equation, is no longer valid (Zhang 1998).  At the quasi-steady state (QSS)
of the wave mode, it can be shown that in fact 
$T_1(R) = d(C_a+C_g)/dR \equiv dC/dR$
(Zhang 1999)!  An intuitive derivation of this equality can be given as
follows: $ d(C_a+C_g)/dR \equiv dC/dR$ is the wave angular momentum flux gradient
in the Eulerian picture, and $T_1(R)$ is the rate of angular momentum
loss for the disk matter in a unit-width annulus located at R in the
Lagrangian picture.  At the quasi-steady
state of the wave mode, these two need to balance each other so the wave
amplitude does not continue to grow (i.e., all the negative angular momentum 
deposited by the wave goes to the basic state of the disk matter 
and none goes to the wave itself,
so that the wave amplitude does not continue to grow, as required by the
condition for the quasi-steady state of the wave mode).

The past calculations of the secular angular momentum redistribution
rate (i.e. Gnedin et al. [1995], Foyle et al. [2010]) considered only
the contribution from gravitational torque couple and ignored the contribution
of the advective torque couple (which cannot be directly estimated using the
observation data, except through our round-about way of estimating the
total torque using the volume-type of torque integral $T_1(R)$).
In the following, we will show that the advective contribution
to the total torque in fact is several times larger than the
contribution of the gravitational torques in the extreme nonlinear regime
usually encountered in observed galaxies, and is of the same sense
of angular momentum transport as the gravitational torque couple -- another
characteristic unique to the nonlinear modal case.  Furthermore, the
two-humped shape of the volume torque distribution (with zero crossing at CR), 
which is equivalent to the two-humped distribution of the phase shift,
is consistent with the bell-shaped torque couplings previously
found in both N-body simulations (Zhang 1998)
and in observed galaxies (Gnedin et al. 1995; for the gravitational torque
coupling contribution only).  This
characteristic distribution is another important piece of evidence
that the density waves present in disk galaxies are in fact spontaneous
unstable modes in the underlying basic state of the disks.

\begin{figure}[ht]
\bigskip
\includegraphics[height=2.5in, width=2.8in, angle=0]{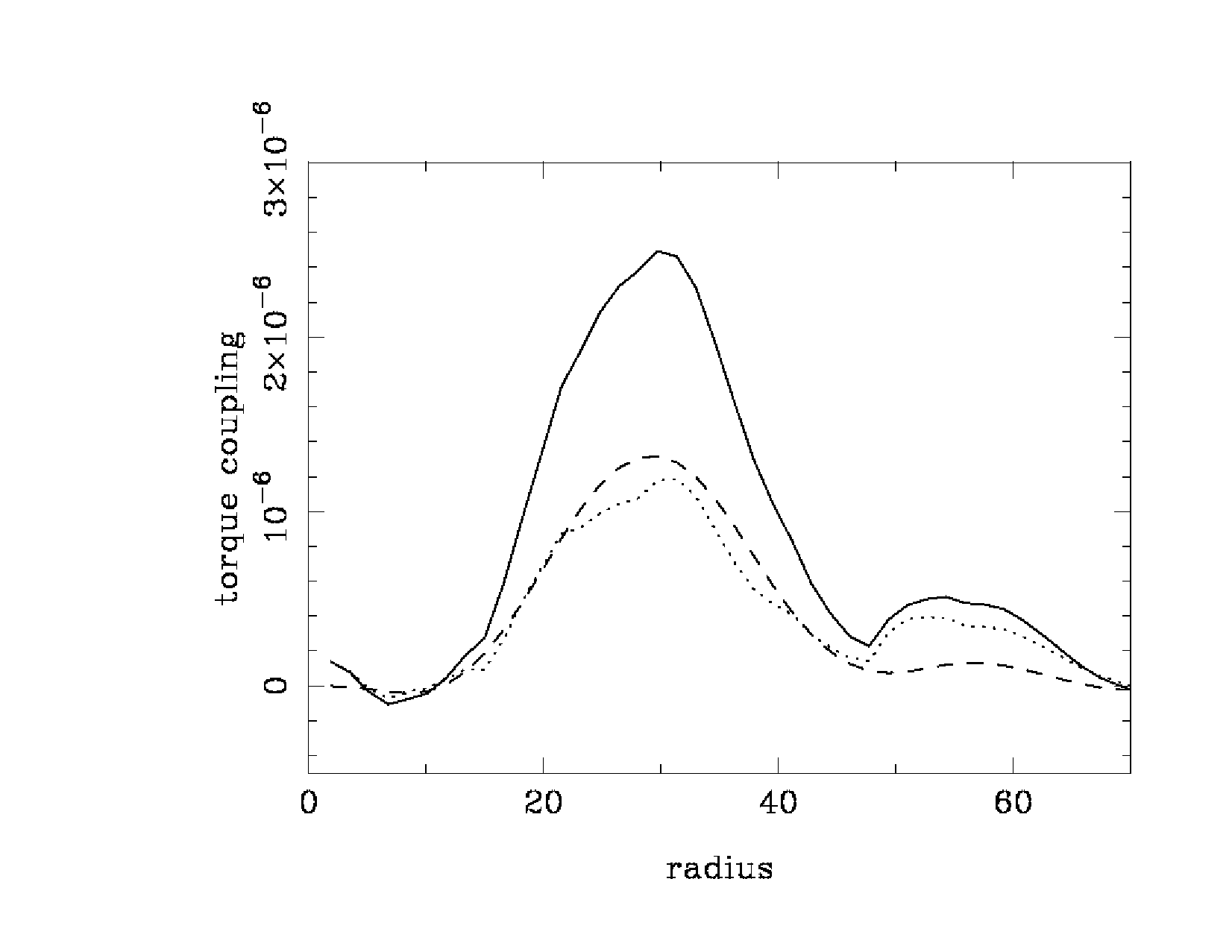}
\includegraphics[height=2.2in, width=2.4in,angle=0]{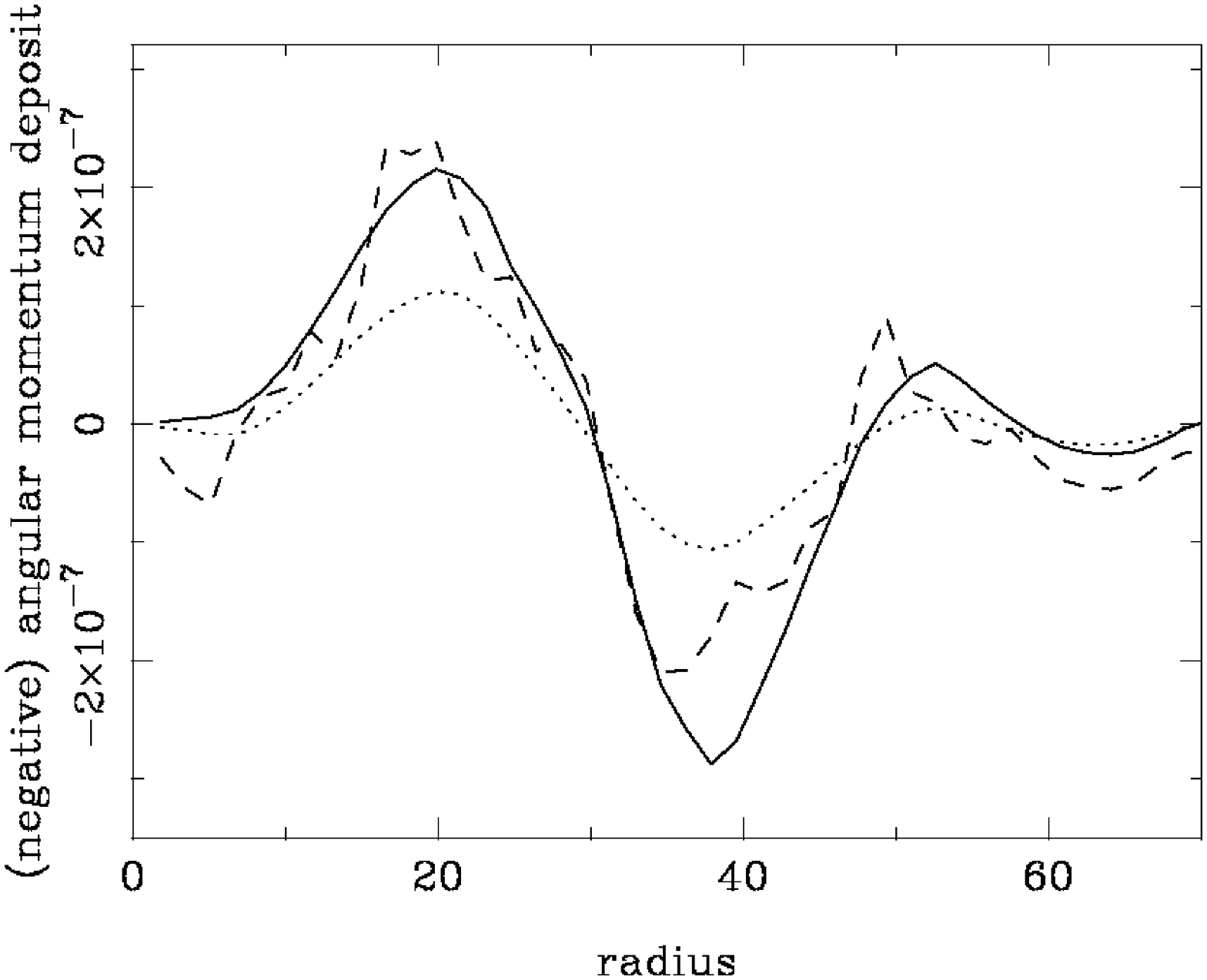}
\caption{{\it Left}: Gravitational (dotted), advective (dashed), 
and total (solid) torque couples
from the N-body simulations of Zhang (1996,1998).
{\it Right}: Gradient of gravitational (dotted) and total (dashed) torque couples,
and the volume torque $T_1$ (solid), from the same N-body simulations
(previously unpublished).}
\label{fg:fg1}
\end{figure}

In Figure \ref{fg:fg1}, left, we show the result of N-body calculated gravitational
and advective torque couplings obtained in Zhang (1996,1998), which shows obviously
the bell-shaped curves for both type of couples, and with the peak near the
CR of the dominant spiral mode at r=30.
In Figure \ref{fg:fg1}, right, we show the gradient of the gravitational and
total torque couples, and compare them with $T_1(R)$.  
It is clear that $T_1(R) \neq
dC_g/dR$, and rather is closer to $dC/dR$, though the equality is not yet
exact because this particular simulated N-body mode never achieved true steady
state. The second hump in the left plot is due to a spurious edge mode.
In physical galaxies, as our examples below will show, the difference between
$T_1(R)$ and $dC_g/dR$ becomes even more pronounced than in these N-body
simulations because of the higher degree of nonlinearity of the wave modes
in physical galaxies.

\section{Examples of Phase-Shift, Volume Torque,
and Mass-Flow Analysis for Individual Galaxies}

\subsection{NGC 4321 (M100)}

\begin{figure}[ht]
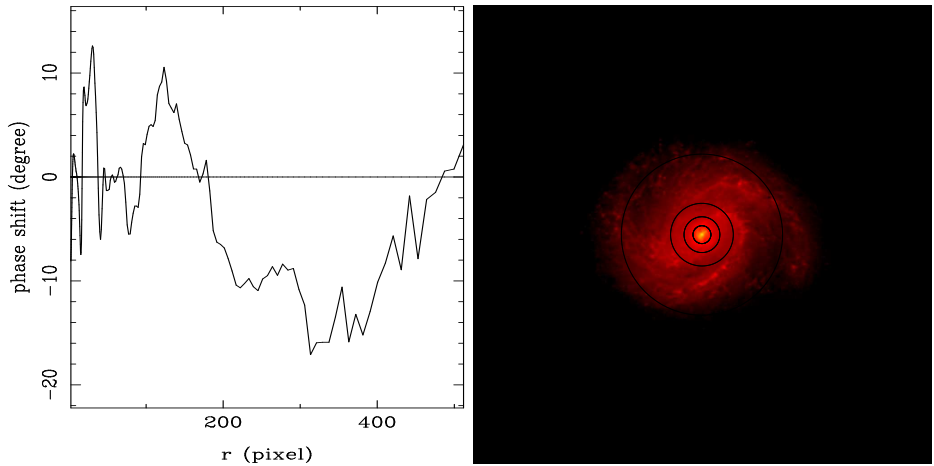

\bigskip
\includegraphics[height=2.4in, width=2.4in, angle=0]{phaseshiftn4321.ps}
\includegraphics[height=2.4in, width=2.4in,angle=0]{overlayn4321.ps}
\caption{{\it Left}: Calculated phase shift versus galaxy radii for NGC 4321.
Two corotation radii are indicated, as is the location of the ends of the
bar. {\it Right}: Deprojected mid-infrared $3.6 \mu m$ SINGS image 
of NGC 4321 in log scale, with the corotations determined by the phase shift 
method superimposed as circles. From Zhang \& Buta (2007).}
\label{fg:fg4321}
\end{figure}

In Figure \ref{fg:fg4321} we plot the phase shift vs galactic radii 
for late-type barred galaxy NGC 4321 (M100),
as well as the CRs determined by the major positive-to-negative (P/N) crossings
and overlaid on the galaxy image (Spitzer $3.6 \mu m$ SINGS survey, Kennicutt et al. 2003).
used for carrying out this analysis.
We find four well-resolved corotation radii for this galaxy (Zhang \& Buta 2007).

\begin{figure}[ht!]
\vspace{3.9in}
\includegraphics{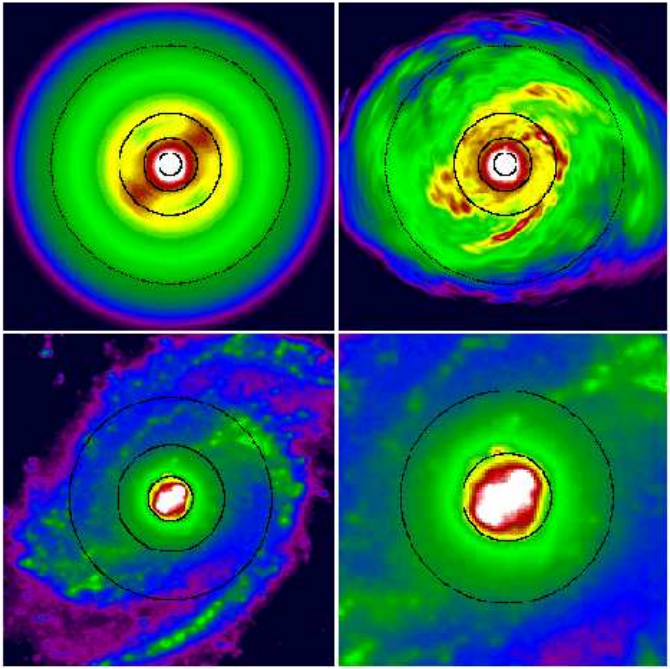}
\caption{{\it Top Left:} Bar-separated
SINGS image of the bright inner region of NGC 4321 superimposed
with the 4 corotation circles determined using the phase shift method.
The box size of the image is 6\rlap{.}$^{\prime}$4 by
6\rlap{.}$^{\prime}$4.
{\it Top Right:} Spiral-separated
SINGS image of NGC 4321 of the same region as at left superimposed
with the 4 corotation circles determined using the phase shift method.
The box size of this image is also 6\rlap{.}$^{\prime}$4 by
6\rlap{.}$^{\prime}$4.
{\it Bottom Left:} SINGS image (without bar-spiral separation) of
NGC 4321 with a factor of 2 linear
zoom compared to the top panels (box size 3\rlap{.}$^{\prime}$2 by
3\rlap{.}$^{\prime}$2),
superimposed with the central 3 corotation
circles determined using the phase shift method.
{\it Bottom Right:} SINGS image (without bar-spiral
separation) of NGC 4321 with a factor of 4 linear
zoom compared to the top panels (box size 1\rlap{.}$^{\prime}$6 by
1\rlap{.}$^{\prime}$6),
superimposed with the central 2 corotation
circles determined using the phase shift method. From Zhang \& Buta (2007).}
\label{fg:fgzoom}
\end{figure}

In Zhang \& Buta (2007), we have used the SINGS image to perform a
bar-spiral separation using the methods described by Buta et al.
(2005), the results of which are shown with our four CR circles
superposed in the top two panels of
Figure \ref{fg:fgzoom}.  The lower two frames of Figure
\ref{fg:fgzoom} present the original
image (without bar-spiral separation) zoomed-in by a linear factor
of 2 and 4, respectively, compared to the top two frames.
Here we see clearly that the innermost corotation circle (CR$_1$) encloses
the strong secondary bar.  Between the next two CRs
(CR$_2$ and CR$_3$) there appear to be faint spiral structures.

\begin{figure}[ht]
\bigskip
\begin{center}
\includegraphics[height=2.4in, width=2.4in, angle=0]{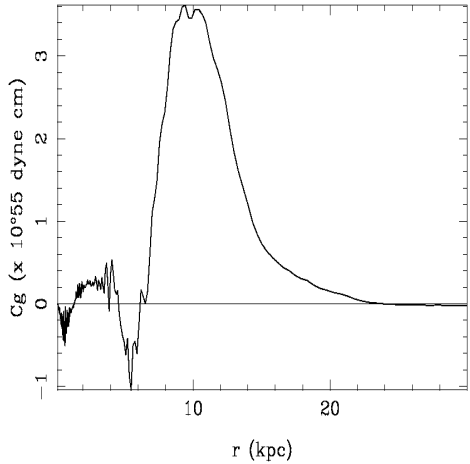}
\end{center}
\caption{Calculated gravitational torque coupling versus radius for galaxy NGC 4321.}
\label{fg:fg4321torque}
\end{figure}

In Figure \ref{fg:fg4321torque}, we plot the calculated gravitational
torque couple for NGC 4321 using the SINGS 3.6 $\mu m$ image.  This
torque calculation result is very similar in shape to the one calculated
for the same galaxy by Gnedin et al. (1995), though the scale factor
is more than a factor of 10 smaller than obtained in their paper.
We have tried to switch to use an R-band image as in Gnedin et al.,
and rescaled the galaxy parameters to be in agreement 
with what they used, still the resulting scale 
is smaller by a factor of 5 from that in the Gnedin et al. (1995).
This same amount of scaling difference is recently found by Foyle et al. (2010)
as well, when they try to reproduce the Gnedin et al. result.  
Therefore it is possible that the
Gnedin et al. calculation suffered an internal error somewhere, since our
result and the Foyle et al. result were obtained entirely independently
and using data from different passbands.

\begin{figure}[ht]
\bigskip
\includegraphics[height=2.4in, width=2.4in, angle=0]{accretion4321.ps}
\includegraphics[height=2.4in, width=2.4in,angle=0]{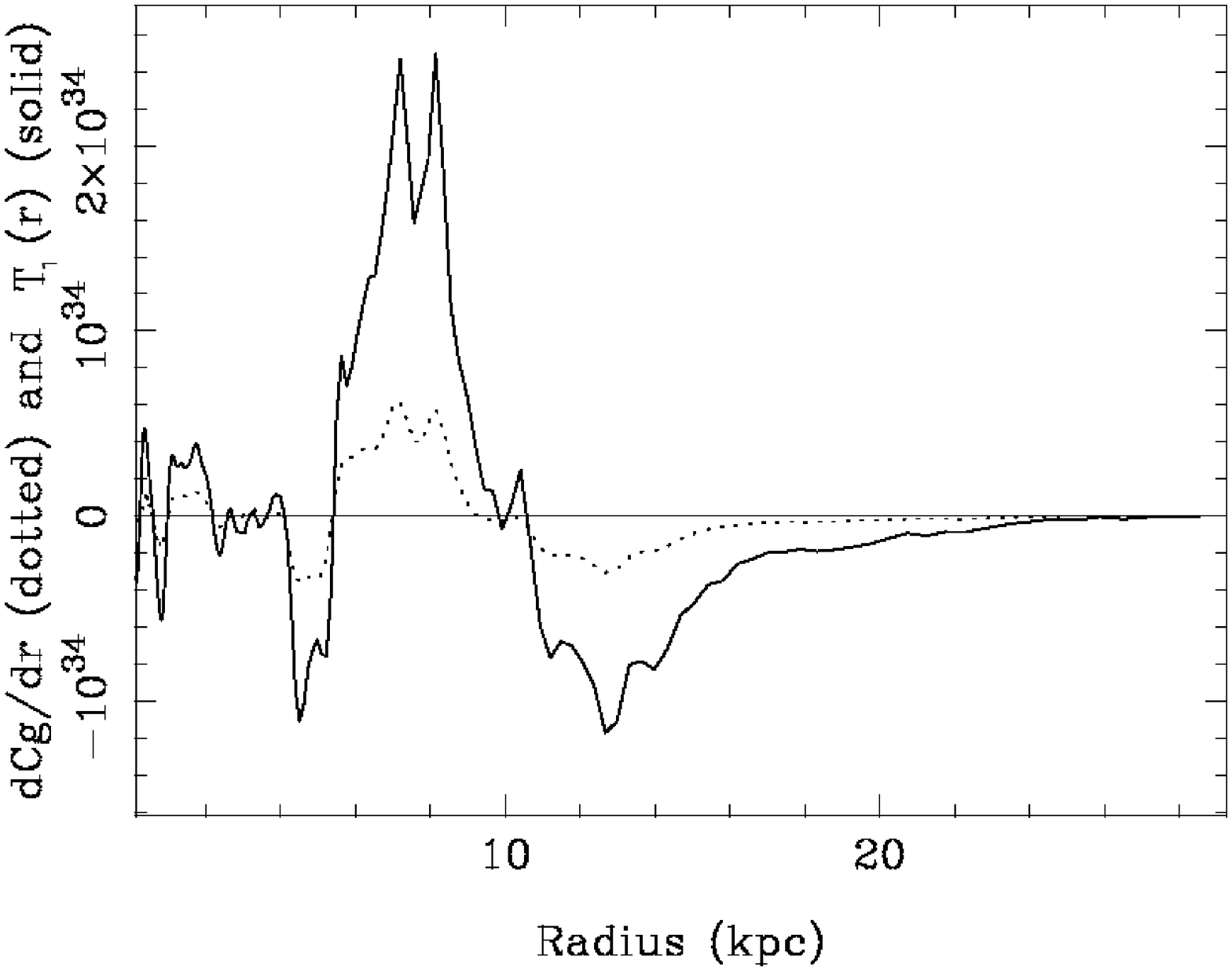}
\caption{{\it Left}: Calculated stellar mass flow rate radius for galaxy NGC 4321.
{\it Right}: Calculated radial gradient of gravitational torque 
coupling integral compared
with the volume torque integral for galaxy NGC 4321 (unit: dyne).}
\label{fg:fg4321a}
\end{figure}

In Figure \ref{fg:fg4321a}, we show the calculated mass flow rates and
radial gradient of gravitational torque coupling integral $C_g(R)$ as compared to the
volume torque integral $T_1(R)$.  
There is about a factor of 4 difference between the
volume torque integral and the gradient of the gravitational torque integral,
indicating that the remainder, which is contributed by the advective torque
coupling, is in the same sense, but much greater in value, than the gravitational
torque coupling integral.  Note that this difference between the volume
torque integral and the gradient of the (surface) gravitational torque couple
is only expected in the new theory: in the traditional theory of LBK these two are supposed
to be equal to each other.  Furthermore, if the LBK theory is used literally,
one should not expect any mass flow rate at all expect at isolated resonance
locations (since the total angular momentum flux is expected to be constant).
The existence of a mass flux across the entire galactic disk is also
contrary to the LBK's original expectations.

\subsection{NGC 5194 (M51)}

\begin{figure}[ht]
\smallskip
\includegraphics[height=2.4in, width=2.4in, angle=0]{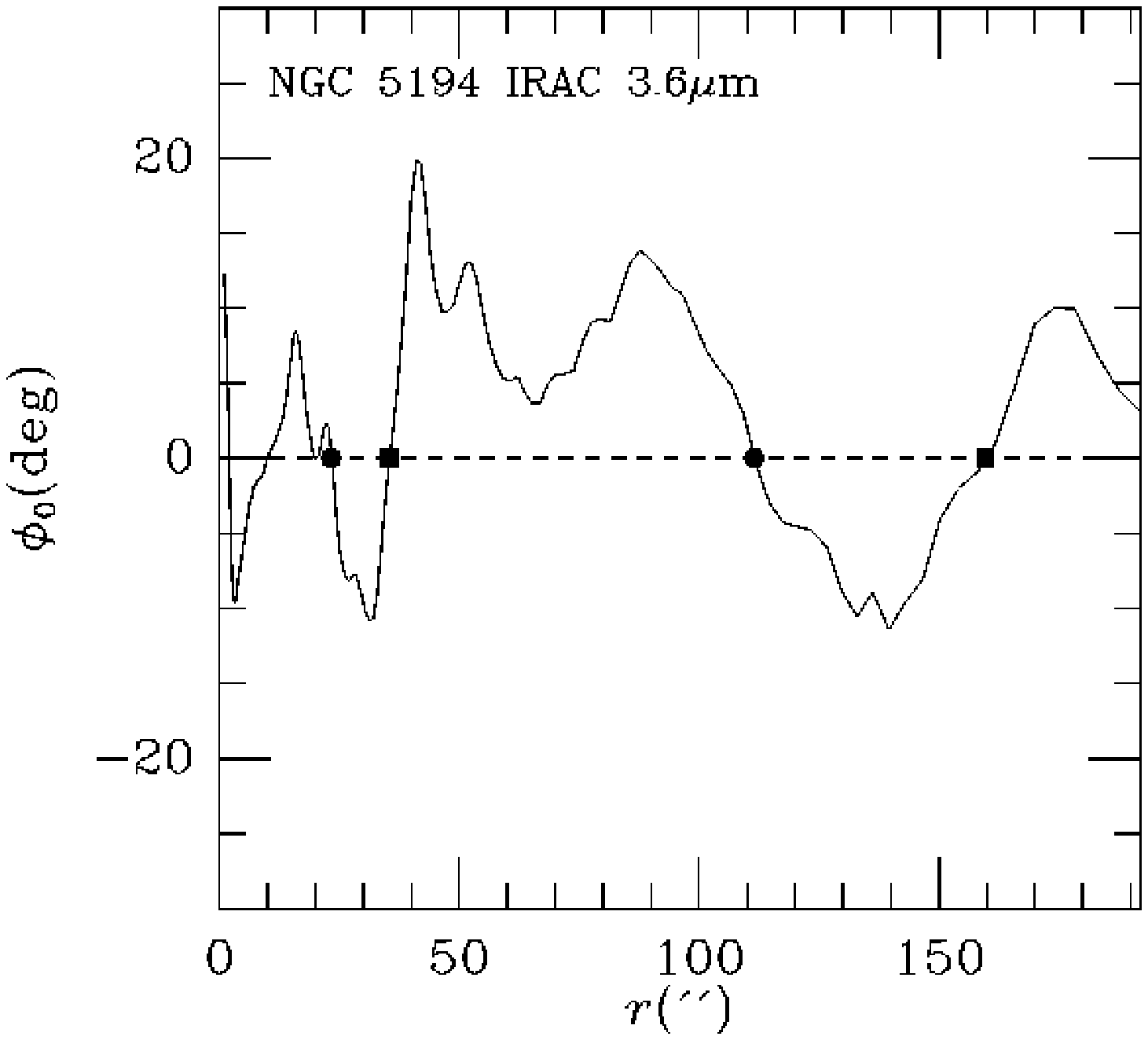}
\includegraphics[height=2.4in, width=2.4in,angle=0]{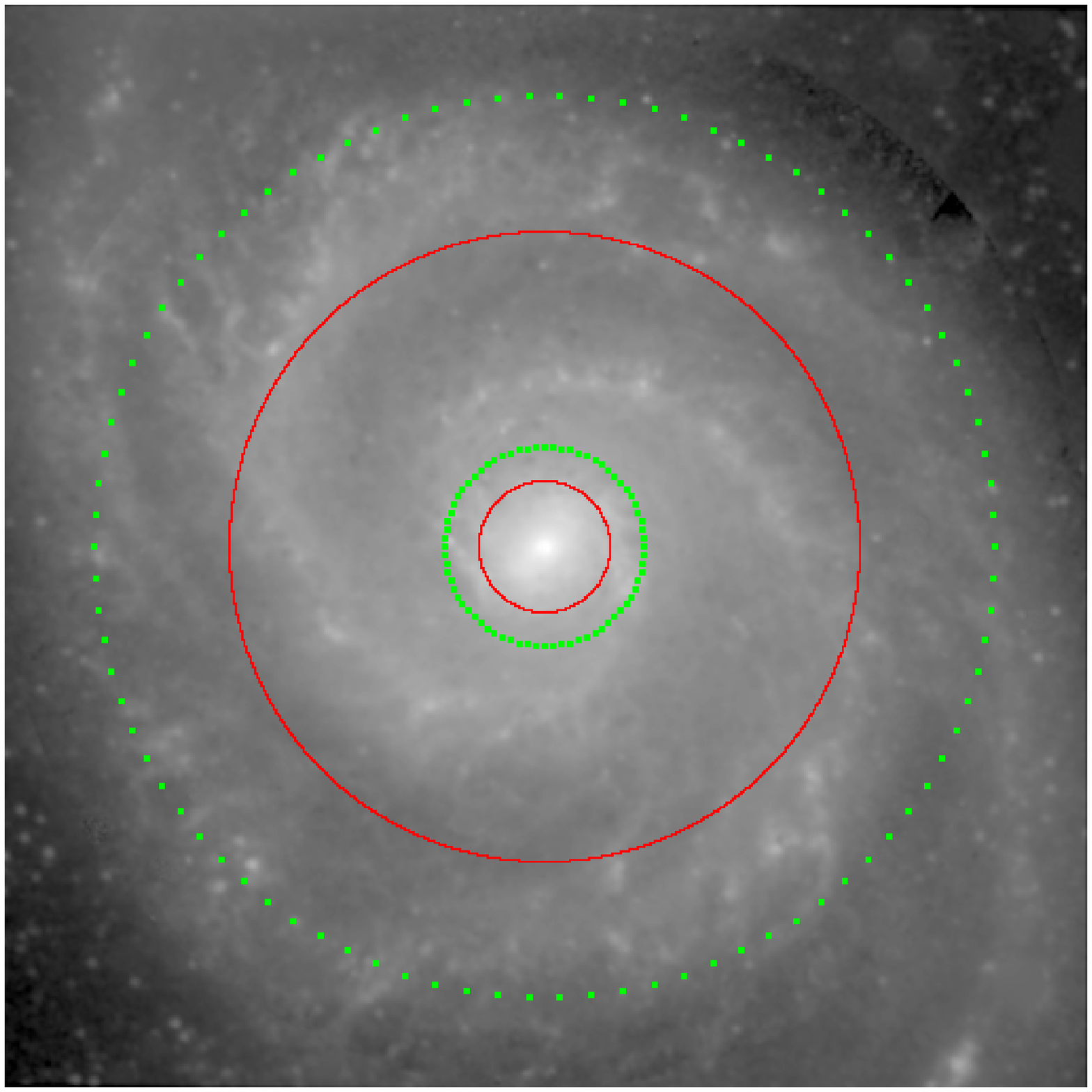}
\caption{{\it Left}: Calculated phase shift vs galaxy radii for NGC 5194
(M51).  Two corotation radii are indicated, as well as the mode decoupling
points.  {\it Right}: Deprojected mid-infrared $3.6 \mu m$ SINGS image 
of NGC 5194 (M51), with the CRs determined by the phase shift 
method superimposed as solid red circles, and mode decoupling radii as
dashed green circles.}
\label{ngc5194}
\end{figure}

We have also used the potential-density phase shift method on the 
interacting galaxy NGC 5194 (M51)
(Figure~\ref{ngc5194}). By focusing on the area that just excludes 
the small companion NGC 5195, which is likely to lie outside of the M51
galactic plane and thus have minor influence on the internal dynamics of M51 
at the epoch of observation (a conjecture which is confirmed by our analysis),  
the phase shift analysis gives two major CR radii (P/N crossings on the
phase shift plot, represented by red circles on the overlay image)
followed by two negative-to-positive (N/P) crossing radii, represented
by the green circles on the image; the latter
are believed to be where the inner mode decouples from the outer mode.
These radii match very well the galaxy morphological features (i.e.,
the inner CR circle lies near the end of the bar, and
the first N/P crossing circle is where the two modes are seen
to decouple). For the outer mode, the CR circle seems to just
bisect the regions where the star-formation clumps are either
concentrated on the inner edge of the arm, or on the outer edge of the
arm -- a strong indication that this second CR is located right near where
the pattern speed of the wave and the angular speed of the stars match
each other. This supports the hypothesis that the spiral patterns in
this galaxy are intrinsic modes rather than tidal transients,
and that tidal perturbation serves to enhance the
prominence of the intrinsic mode, but does not alter its modal shape.

\begin{figure}[ht]
\bigskip
\includegraphics[height=2.4in, width=2.4in, angle=0]{accretion5194.ps}
\includegraphics[height=2.4in, width=2.4in,angle=0]{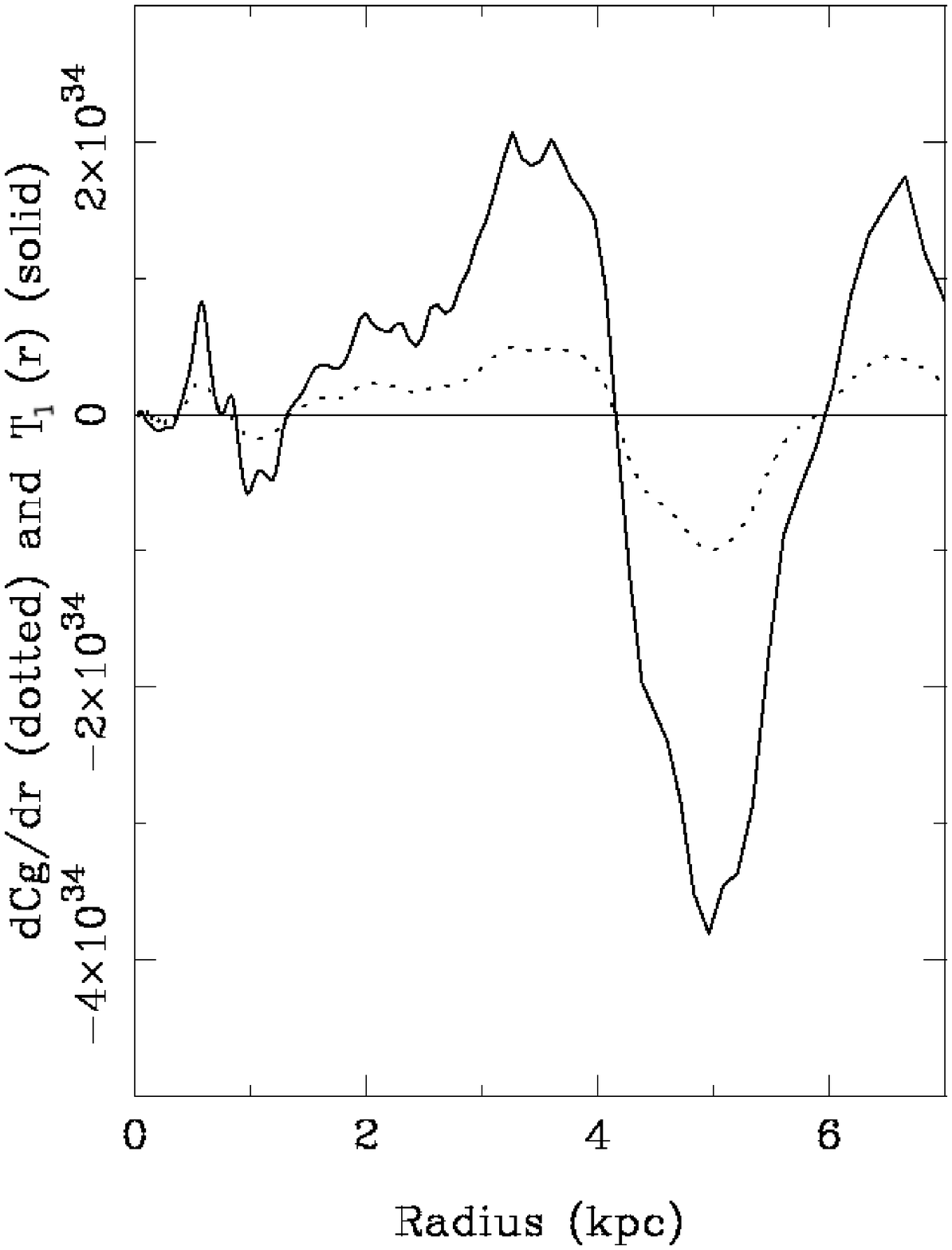}
\caption{{\it Left}: Calculated stellar mass flow rate vs radius, and 
{\it Right}: Calculated radial gradient of torque coupling integral compared
with the volume torque integral, for galaxy NGC 5194 (unit: dyne). }
\label{fg:fg5194a}
\end{figure}

In Figure \ref{fg:fg5194a}, we plot the calculated mass flow rates and
the comparison of the gradient of gravitational torque couple $C_g(R)$
and volume torque $T_1(R)$ for NGC 5194.  Once again we see that for this
galaxy, as for NGC 4321, the advective torque couple is in the same sense
as the gravitational torque couple,
and dominates the value of the total torque coupling.
Similar results were found for other galaxies we have analyzed so far as well.

\section{Conclusions}

The correct treatment of gravitational many-body systems containing self-organized
global patterns, such as density wave modes in disk galaxies, requires a re-examination
of classical dynamical approaches and assumptions.  Our experience so far has shown
that entirely new qualitative and quantitative results can emerge from the
collective interactions of the many particles in a complex dynamical system.
Formerly sacred laws (such as the differential form of the Poisson equation) can
break down at the crest of collisionless shocks, and new meta-laws (such as the
equality of the volume torque integral with the derivative of the sum of gravitational
and advective surface torque coupling integrals) appear as emergent laws.
Such emergent behavior is the low-energy Newtonian dynamical analogy of 
high energy physics' spontaneous breaking of gauge symmetry, a well known
pathway for forming new meta laws when traversing the hierarchy of organizations.

\section{References}

Binney, J., \& Tremaine, S. 2008, Galactic Dynamics, second ed. (Princeton:
Princeton Univ. Press)

Buta, R.J., Vasylyev, S., Salo, H., Laurikainen, E., 2005, AJ, 130, 506

Buta, R.J., \& Zhang, X. 2009, ApJS, 182, 559

Foyle, K., Rix, H.-W., \& Zibetti, S. 2010, MNRAS, 407, 163

Gnedin, O., Goodman, J., \& Frei, Z. 1995, AJ, 110, 1105

Haan, S. et al. 2009, ApJ, 692, 1623

Kennicutt, R.C. Jr., et al. 2003, PASP, 115, 928

Kormendy, J., \& Kennicutt, R.C. 2004, ARAA, 42, 603

Lynden-Bell, D., \& Kalnajs, A.J. 1972, MNRAS, 157, 1

Sellwood, J.A., 2010, arXiv1008.2737

Zhang, X. 1996, ApJ, 457, 125; 1998, ApJ, 499, 93; 1999, ApJ, 518, 613

Zhang, X., \& Buta, R.J., 2007, AJ 133, 2584
\end{document}